\documentclass[acmsmall]{acmart}

\setcopyright{none}
\renewcommand\footnotetextcopyrightpermission[1]{}
\copyrightyear{2018}

\acmYear{2018}
\acmDOI{10.1145/1122445.1122456}

\acmBooktitle{Woodstock '18: ACM Symposium on Neural Gaze Detection,
  June 03--05, 2018, Woodstock, NY}
\acmPrice{15.00}
\acmISBN{978-1-4503-XXXX-X/18/06}

\begin{document}

%%
%% The "title" command has an optional parameter,
%% allowing the author to define a "short title" to be used in page headers.
\title{SurviveCovid-19++ : A collaborative healthcare game towards educating people about safety measures and vaccination for Covid-19}

\author{Akhila Sri Manasa Venigalla} 
\email{cs19d504@iittp.ac.in}
\author{Dheeraj Vagavolu}
\email{cs17b028@iittp.ac.in}
\author{Sridhar Chimalakonda}
\email{ch@iittp.ac.in}
\affiliation{
\institution{\\
Research in Intelligent Software \& Human Analytics (RISHA) Lab,
\\Department of Computer Science and Engineering, \\
Indian Institute of Technology Tirupati}
 \city{Tirupati}
 \country{India.}
}
\renewcommand{\shortauthors}{Venigalla et al.}

%%
%% The abstract is a short summary of the work to be presented in the
%% article.
\begin{abstract}
 \textit{Covid-19} has been affecting population across the world for more than an year, with diverse strains of this virus being identified in many countries. 
Vaccines to help in curbing the virus are being developed and administered. Preventing the spread of the disease requires collaborative efforts from everyone. People with varied professional backgrounds have varied responsibilities in controlling the pandemic. It is important that everyone is aware of their respective responsibilities and also empathize with efforts and duties of other individuals. It is here, we wish to leverage the potential of games in healthcare domain, towards educating about \textit{Covid-19}.
  With an aim to educate the population about vaccination against \textit{Covid-19}, responsibilities of citizens with varied professional backgrounds, and emphasize on the need for collaboration to fight against the pandemic, by following safety measures, we present \textit{SurviveCovid-19++}, a collaborative multiplayer desktop based game. The game essentially revolves around four roles -  \textit{doctor}, \textit{sanitation worker}, \textit{citizen} and \textit{law enforcer}, delivering their duties, following safety measures and collaboratively clearing multiple stages in the game. 
%   This game is an extended version of SurviveCovid-19, developed towards educating people about individual safety measures that are to be followed against \textit{Covid-19}. 
 We have performed a preliminary evaluation of the game through a qualitative and quantitative user survey. The results of the user survey were encouraging, with volunteers expressing their increased empathy towards efforts of individuals with varied professional backgrounds, and better understanding of the importance of safety measures against \textit{Covid-19}.  
\end{abstract}

%%
%% The code below is generated by the tool at http://dl.acm.org/ccs.cfm.
%% Please copy and paste the code instead of the example below.
%%

%%
%% Keywords. The author(s) should pick words that accurately describe
%% the work being presented. Separate the keywords with commas.

%%
%% This command processes the author and affiliation and title
%% information and builds the first part of the formatted document.
\maketitle

\section{Introduction}
The Covid-19 pandemic has largely affected the day-to-day lives of many \cite{haleem2020effects} and also had a huge impact on the economy of several countries \cite{altig2020economic}. While the first wave of the pandemic seemed to settle down during the summer of 2020, there started to be a rise in the number of cases across the world due to the second wave of the pandemic in the fall of 2020 \cite{ioannidis2021second}. Newer strains of the virus are being identified in various countries during the second wave, which also resulted in rapid spread and more serious impact on the patients' health, than the basic strain \cite{le2021alert, seligmann2020inverted, de2021second}. There is a significant rise in the number of cases, affecting even the younger population, across many countries that witnessing the second wave \cite{le2021alert, seligmann2020inverted, de2021second}. Around 2.84 million deaths have been reported across the world due to Covid-19, as of 6 April, 2021\footnote{\url{https://www.who.int/publications/m/item/weekly-epidemiological-update-on-covid-19---6-april-2021}}. On 5 April, 2021, India alone recorded 103K cases\footnote{\url{https://www.who.int/india/emergencies/coronavirus-disease-(covid-19)/india-situation-report}}, which is the all time highest number of cases in 24 hours span, since the Covid-19 outbreak in the country, and notable across the world according to WHO report\textsuperscript{1}. 

Vaccines are being developed to control the spread of the virus and are being proactively administered to people across the world. Social welfare campaigns and announcements are being conducted by various governments across the world to improve awareness among the people to get vaccinated. For example, the Government of India has started vaccine communication strategy, that included setting up telephone lines to resolve doubts and misconceptions of the general public towards the vaccine, advertisements in press and national media, and so on\footnote{\url{https://www.mohfw.gov.in/pdf/Covid19CommunicationStrategy2020.pdf}}. Such strategies emphasize the need to educate people about the importance of taking the vaccines and its uses. 

With around eight strains of the virus identified across the world\footnote{\url{https://www.cdc.gov/coronavirus/2019-ncov/cases-updates/variant-surveillance/variant-info.html\#Consequence}}, scientists are working towards developing vaccines to control each of these strains \cite{forni2021covid, lurie2020developing}. The sanitation staff in majority of the countries are carrying out sanitization works in public places to create hygenic conditions, which could reduce the spread of the virus \cite{khan2020sanitization, cheng2020safely}. Doctors, nurses and medical students across the world have been performing their duties impeccably, almost round the clock, to cure the patients effected by Covid-19 \cite{kumaraiah2020innovative, miller2020role, mcconnell2020balancing, tan2020burnout}. 
With safety measures such as usage of masks and limited strength in public gatherings made mandatory in many countries, common people in across the world are also expected to follow the safety precautions and fulfill their duty of controlling the pandemic. Considering these varied duties and responsibilities of people with varied professional backgrounds, collaborative efforts and mutual support among all professions is essential to control the spread of the virus \cite{chakraborty2020extensive, wen2020many}. 
This also emphasizes the need for individuals to understand the importance of each profession to control the pandemic and empathize with each other.

% \footnote{\url{https://www.who.int/publications/m/item/weekly-epidemiological-update-on-covid-19---6-april-2021}}. 

% Covid-19 affects many. 

% Varied strains of the virus - rapid spread and dangerous

% Vaccines being developed

% Need to educate people on importance  and uses of taking vaccines.

% Need to bring out empathy for evryone fighting against the pandemic in their own ways. 
% Need for people to know the importance of each profession and the need to collaboratively work towards controlling the pandemic.

% Vaccine for one strain maybe less effective for other. 

% Need to understand this, and consequently develop vaccines.

% Nurses to be aware of different vaccines for different strains

Games support effective learning and contribute to improved retention capacity \cite{habgood2011motivating, de2018games}. 
Several games have been developed to support learning and teaching various aspects, which include programming, mathematics, languages, health care measures, and so on \cite{de2018games, sharifzadeh2020health}. It has been observed that games that involve collaborative aspects increase user engagement and consequently support better and motivated learning \cite{bigdeli2017digital}. Collaborative aspects also improve communication in a team and thus support in enhancing the team spirit, while learning new concepts in the domain of interest of the game \cite{buchinger2018guidelines}. Such games have been introduced in various fields such as science and medicine and were observed to enhance the learning experience of users \cite{sung2013collaborative, hannig2012emedoffice}.

Though there are many educational and motivational games, there exist very few games to bring awareness about Covid-19 and the safety measures to be followed against Covid-19. SurviveCovid-19 \cite{venigalla2020survivecovid} has been developed to bring awareness on the health measures to be followed to control Covid-19. However, it only supports single player version, and is only from a citizen's perspective. Considering the importance of collaborative effort required to control the pandemic, the need to educate people about importance and uses of vaccines, and the advantages of collaborative games, a game that supports collaborative, multi-player version could better contribute in educating the users. Also, the existing SurviveCovid-19 game does not include any information about the vaccine or strains of the virus. With vaccination being administered and with growing strains of the virus, educating people about the differences in these strains and the vaccines could be more useful. 

Hence, in this paper, we present \textit{\textbf{SurviveCovid-19++}, a multi-player, collaborative game, that helps in understanding collaborative effort required by people with different professional backgrounds such as doctors, law enforcers and so on, in controlling the pandemic, while including the aspects of safety measures, administering vaccines and multiple strains of the virus}. \textit{SurviveCovid-19++} supports four individuals with varied professions - \textit{Doctor}, \textit{Citizen}, \textit{Sanitation Worker} and \textit{Law Enforcer}. It facilitates doctors to treat people affected with \textit{Covid-19} in the game through medicines, law enforcers to break crowds in the city, citizens to pick up groceries and sanitation workers to sanitize the localities and kill the virus. It also encourages players of all professions to use masks and sanitizers to prevent themselves from being effected, while delivering their duties, and also requires all the players to work collectively towards collecting vaccine parts in each stage.
We perform a preliminary evaluation on the influence of \textit{SurviveCovid-19++} on players' understanding of individual and collaborative efforts, and safety measures towards controlling the spread of \textit{Covid-19} based on factors determining pedagogical value. We also evaluate the player experience and usability  through a quantitative user survey, based on  and MEEGA+ evaluation model.

\section{Related Work}
% Importance of Collaborative learning
Considering the playfulness and motivation inducing factor of games, they have been adapted in the healthcare domain to treat patients, to educate individuals about healthcare practices and motivate them to follow these practices, and also to train healthcare students \cite{habgood2011motivating,costa2015systematic, graafland2018serious}. 

\subsection{Games for Healthcare}
Nintendo Wii-Fit videogame system has been used to facilitate in-home exergaming, which included exercise through games, to treat patients suffering from rheumatoid arthritis \cite{ambrosino2020exergaming}. A controlled study of using this system has resulted in positive outcomes, indicating exergaming to be an effective rehabilitation tool \cite{ambrosino2020exergaming}. Games with racing and matching rhythm themes were used as a part of training in prosthetic motor rehabilitation and were observed to encourage the patients to continue the training, than dropping out of the process, as in the case of traditional rehabilitation \cite{prahm2018playbionic}. 

Hochsmann et al. have developed a game to promote physical activity and exercise among diabetic patients \cite{hochsmann2019novel}. A study on impact of this game revealed increased motivation among the participants towards physical exercise, which resulted in controlled glucose levels \cite{hochsmann2019novel}. \textit{Alien Health Game} has been designed to teach nutritional values to children and to educate them on healthy food choices \cite{hermans2018feed}. \textit{Monster Apetite }game has been developed to encourage healthy lifestyles and consequently control obesity \cite{hwang2017monster}. This game displays a visual image of monster based on players' in-game food choices, thus providing a perception on the nutritional value of food choices \cite{hwang2017monster}. Bomfim et al. have proposed \textit{Pirate Bri's Grocery Adventure} to improve food literacy among individuals that could help them in making healthy choices while purchasing food items. It has been observed that this game could enable users in controlling impulse purchases, and also improved their understanding on nutritional values of food items \cite{bomfim2020food}. \textit{MicroQuest} has been developed to educate children about microbiology and bring awareness among them on healthcare issues that could be caused due to unhygienic conditions \cite{molnar2018learning}. It facilitates players to travel through the human body and other places such as kitchen  and enables them to explore presence of microbes, bacteria and viruses \cite{molnar2018learning}. 

% Games for healthcare
\subsection{Collaborative Games for Healthcare}
Collaborative and competitive games were observed to increase motivation among individuals to play educational games, which further improves their understanding on the topics being taught \cite{al2014starsrace, johnson2017creating}. \textit{StarsRace} has been developed as a collaborative mobile game with an aim to control obesity by requiring players to collect stars virtually, which would be placed at specific geographical locations \cite{al2014starsrace}. This enables players to move around the map physically and gain points \cite{al2014starsrace}. \textit{LIVE} has been developed as a virtual environment platform that facilitates learning of self-management skills and provides social support to diabetic patients \cite{johnson2017creating}. It presents a 3D view of various objects in the virtual environment that includes pharmacy items, fitness equipment, avatar choices and so on, while encouraging collaborative games and social networking through out \cite{johnson2017creating}.
\subsection{Games for Pandemics}
Some games, though limited in number, have also been developed to educate and bring awareness among individuals about pandemics and the safety measures to be taken to control the pandemics.  \textit{Pandemic} game has been integrated with artificial intelligence to support players to travel across different places in the world in-game and collaboratively cure cities that get infected by various diseases \cite{sfikas2020collaborative}. \textit{SurviveCovid-19} has been introduced to bring awareness among individuals on the safety measures to be followed to reduce the spread of \textit{Covid-19} \cite{venigalla2020survivecovid}. \textit{COVID-19 - Did You Know?} has been developed as a true-false based question-answering themed game, to educate users about \textit{Covid-19}, by providing information about various topics related to \textit{Covid-19}, such as use of masks, cleanliness, health, social distancing and so on \cite{gaspar2020mobile}. 

Of the limited number of games that consider pandemics, games that support collaborative game play are further less in number, which almost do not exist in the literature with respect to Covid-19. To the best of our knowledge, \textit{Pandemic} is the only collaborative game that considers pandemic aspect. Though controlling \textit{Covid-19} requires collaborative efforts from individuals with varied professional backgrounds, we could not find any collaborative-games in the literature which attempt to educate about \textit{Covid-19}. While \textit{COVID-19 - Did You Know?} \cite{gaspar2020mobile} is a completely question-answering themed game, with minimal gamification elements, \textit{SurviveCovid-19} \cite{venigalla2020survivecovid}, though has reasonable extent of gamification elements, supports only single player version, thus, not supporting collaborative gaming. Further, AI enabled \textit{Pandemic} \cite{sfikas2020collaborative} is not specific to \textit{Covid-19}, and also does not include disease-specific elements or treatments in-game. 

Hence, we present \textit{SurviveCovid-19++}, a collaborative multi-player game, that includes vaccines, masks, sanitizers and pandemic-related game elements, while supporting collaborative game play by including different avatars such as doctor, law enforcer, and so on, within each team of players.
\begin{figure}
    \centering
    \includegraphics[width = \linewidth]{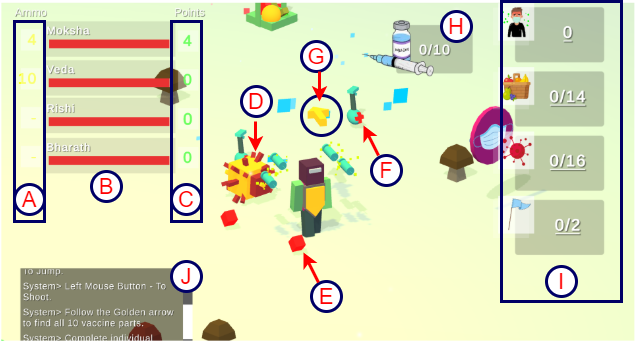}
    \caption{A snapshot of the game being played by four players, depicting [A] count of Ammos currently available for each player, [B] health meter of each player's avatar, [C] health points for each avatar, [D] virus ready to infect player, [E] health vitamin to increase health points when collected, [F] refillable disinfectant, [G] arrow pointing towards the direction of finding vaccine parts, [H] number of vaccine parts collected by all the players, [I] number of items picked up, and target of items to be picked by each avatar in the game and [J] chat box, with list of instructions presented by default.}
    \label{fig:ss}
\end{figure}

% Collaborative games in health-care

% Games for Covid-19

\section{Design of \textit{SurviveCovid-19++}}
Several guidelines for development of serious games for education have been introduced in the literature \cite{de2018gamification, buchinger2018guidelines, wendel2013designing}. Croon et al. have performed a systematic literature review of design guidelines of serious games and collated a list of guidelines based on this review to assist developers of serious games for healthcare \cite{de2018gamification}. These guidelines include selection of gamification mechanics and tailoring the game based on target audience personalities. These guidelines also further incorporate the motivational and self-determination aspects \cite{de2018gamification}. Wendel et al. have proposed design guidelines for collaborative educational games, irrespective of education domain \cite{wendel2013designing}. These guidelines include common goals, communication, heterogenous resources and so on \cite{wendel2013designing}. 
A set of 9 design guidelines for collaborative games in healthcare domain have been proposed by Buchinger et al. based on a systematic literature review of guidelines for collaborative educational games \cite{buchinger2018guidelines}. Features such as intra player interaction, synchronization, rewards and challenge are some of the nine guidelines presented by Buchinger et al \cite{buchinger2018guidelines}. \textit{Sherlock Dengue 8}, a healthcare educational game that aims to teach youngsters about dengue fever, designed based on these guidelines, was observed to help in promoting knowledge acquisition and also increase confidence among the users to control the disease \cite{buchinger2018guidelines}. 

\textit{SurviveCovid-19++} is a multiplayer collaborative game that can be played by a maximum of four players and a minimum of one player. It starts with the storyline of a group of friends in four different professions - \textit{Citizen}, \textit{Doctor}, \textit{Sanitation Worker}, and \textit{Law Enforcer}, in a city, working towards controlling the spread of \textit{Covid-19} in the city and treating the affected individuals. It has defined roles (in-game profession-based) for each avatar and the players can select the avatar of their choice, without repetition in the team. Each avatar is assigned specific tasks, along with collaborative tasks that are to be accomplished to clear each stage of the game. The \textit{Citizen} is required to collect groceries throughout the game. The \textit{Doctor} is required to collect medicine and treat \textit{Covid-19} infected patients with these medicines collected. The \textit{Sanitation worker} needs to collect sanitizer/disinfectant refills throughout the game and use them to disinfect infected locations. The \textit{Law Enforcer} is required to break crowds in the city. All the players are required to collect required number of vaccines apart from their personal goals in the game, while using masks and sanitizers to protect themselves from getting infected by the virus, in the game, to clear each stage. If an avatar gets infected by the virus, the health of the avatar gets reduced and the player's movement in the game is slowed down. The more the player gets infected, slower the movement of the player. A snapshot of the game comprising four players is presented in Fig. \ref{fig:ss}. An arrow is assosciated with each player to show the direction of finding vaccine parts, as shown in Fig. \ref{fig:ss}[G].

\textit{SurviveCovid-19++} has been designed by including the design guidelines for collaborative games, presented in \cite{wendel2013designing} and \cite{buchinger2018guidelines}. 
% and guidelines for serious games in the healthcare context presented in \cite{de2018gamification}
An adaptive set of guidelines, inclusive of all the relevant features from the above mentioned guidelines considered in the design of \textit{SurviveCovid-19++}. These guidelines  and the corresponding adaptation in the game are listed below:

\textbf{Intra-player Interaction \cite{buchinger2018guidelines}/Communication \cite{wendel2013designing} - } All the players in a specific room are provided with a chat application (as shown in Fig. \ref{fig:ss}[J]) that supports interactions among the players in each team.

\textbf{Synchronization \cite{buchinger2018guidelines} - }To support quicker gameplay and to encourage interactions among players, \textit{SurviveCovid-19++} adapts a synchronous gameplay, which allows all the players to simultaneously perform their actions corresponding to their in-game roles (indicated in Fig. \ref{fig:sc1}[D]).

\textbf{Roles \cite{buchinger2018guidelines} - } As bringing out collaboration is an important aim of \textit{SurviveCovid-19++}, different roles are assigned to each player in the game (indicated in Fig. \ref{fig:sc1}[C]). 

\textbf{Resources \cite{buchinger2018guidelines} - } All the players are assigned certain tasks of collecting and using multiple resources, which correspond to various aspects in dealing with \textit{Covid-19}, in the game. For example, \textit{common citizen} in the game has to collect \textit{groceries}, \textit{sanitation worker s} have to collect \textit{disinfectant refills} (indicated in Fig. \ref{fig:ss}[F]), while all the players are required to collect \textit{health vitamins} (indicated in Fig. \ref{fig:ss}[E]) to increase their health score (presented as shown in Fig. \ref{fig:ss}[C]), and \textit{vaccine parts} in each stage, to clear the stage and use these vaccines in the next stages. Also, \textit{masks} and \textit{sanitizers} are to be collected when required in the game, to stay safe from the \textit{virus}. 

\textbf{Score \cite{buchinger2018guidelines} / Scoreboard \cite{wendel2013designing} - } Each player in the game is given a score based on the number of objects and refills collected (as indicated in Fig. \ref{fig:ss}[I] and Fig. \ref{fig:ss}[A] respectively) and the team is also collectively given a score based on the vaccines collected (as indicated in Fig. \ref{fig:ss}[H]). 

\textbf{Challenge \cite{buchinger2018guidelines} - } The virus strain gets stronger in each stage, thus requiring players to collect more number of vaccines. This feature increases the difficulty level of the game, keeps the players informed about the possibility of virus strain growing stronger and the requirement of players to stay alerted by following safety precautions against the virus all the time, while satisfying their roles in the community.

\textbf{Rewards \cite{buchinger2018guidelines} - } The players are also rewarded based on the number of masks and sanitizers collected in the game, thus encouraging them to use masks and sanitizers often, along with collecting various objects to clear the level.

\textbf{Operationalization \cite{buchinger2018guidelines} - } \textit{SurviveCovid-19++} does not require use of any special devices, and thus does not result in \textit{game spread difficulties}. \textit{SurviveCovid-19++} also supports players to take up any of the four roles, thus facilitating unhindered game play, even if there are less than four players.

\textbf{Common goal/Success \cite{wendel2013designing} - } All the players in the game clear each stage together and win or lose the game, through collective effort. Individual players are not facilitated to move to next stages alone, thus enforcing collaboration among the players. 

\textbf{Heterogeneous Resources \cite{wendel2013designing} - } Each avatar in the game has varied abilities and responsibilities. For example, \textit{law enforcer} has the capability to break crowds while the \textit{doctor} does not have. Such features could thus preserve the heterogeneous resources nature of the game.

\textbf{Refillable personal resources \cite{wendel2013designing} - } \textit{SurviveCovid-19++} provides users with meter for health (as indicated in Fig. \ref{fig:ss}[B]), masks and sanitizers. These meters can be refilled when the players visit a doctor, collect masks and sanitizers respectively. These features could thus create a tension in the game among the players and supports collaborative game play.

\textbf{Collectable and tradeable resources \cite{wendel2013designing} - } \textit{SurviveCovid-19++} facilitates players to collect resources specific to their roles and these resources can be traded off when the players wish to swap their roles, to clear a stage. Players can click on `Esc' key to interchange their roles by trading some of their points.

\textbf{Collaborative tasks \cite{wendel2013designing} - } Each player in the game is assigned a specific task, based on the role/avatar taken up by the player. To clear each stage, all the avatars are required to work together and fulfill their roles.

\textbf{Ingame help system \cite{wendel2013designing} - } Instructions and directions are presented at multiple locations through out the game to help players navigate and play the game. This in-game feature has been designed as text on the various objects in the game, so that it does not interrupt the game. 

\textbf{Trading System \cite{wendel2013designing} - } \textit{SurviveCovid-19++} allows players to discuss among themselves if they wish to swap roles to attain collective success. This swapping process requires the players to agree to trade some of their collectibles or points, thus encouraging them to take decisions that could affect the team's win or loss.
\begin{figure}
    \centering
    \includegraphics [width = \linewidth]{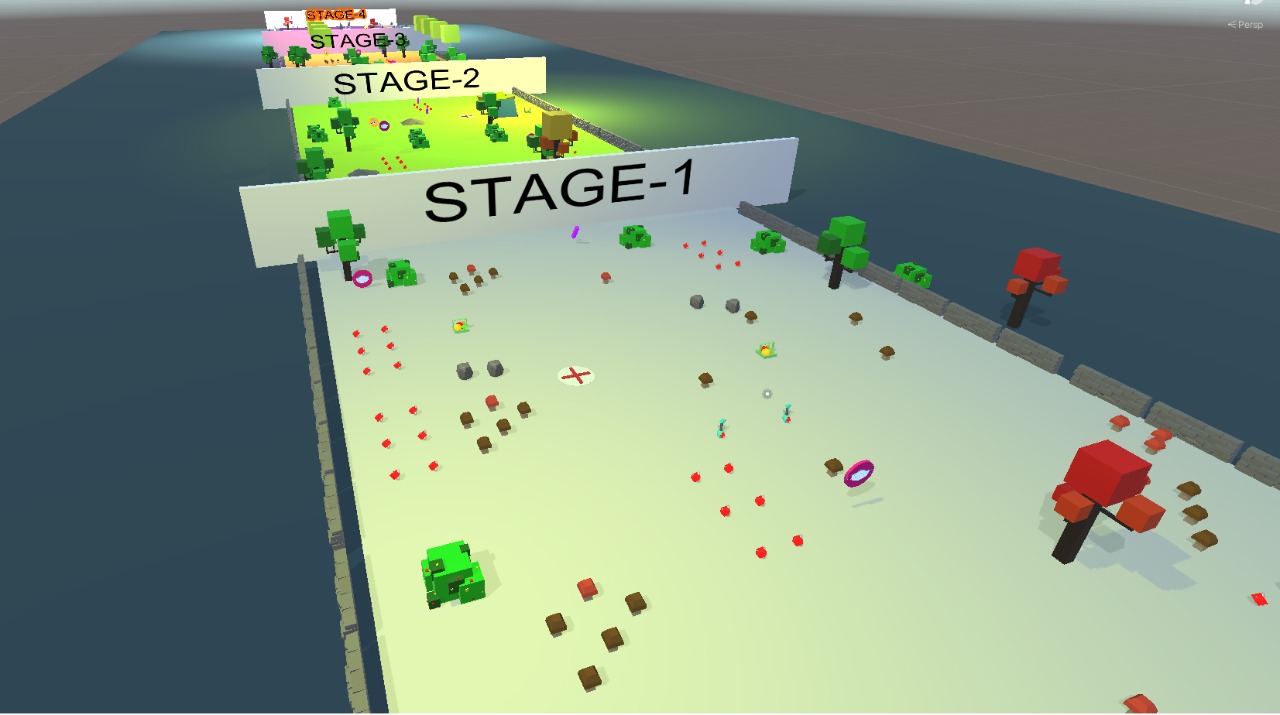}
    \caption{A map indicating the four stages in \textit{SurviveCovid-19++}}
    \label{fig:map}
\end{figure}
\begin{figure}
    \centering
    \includegraphics[width = \linewidth]{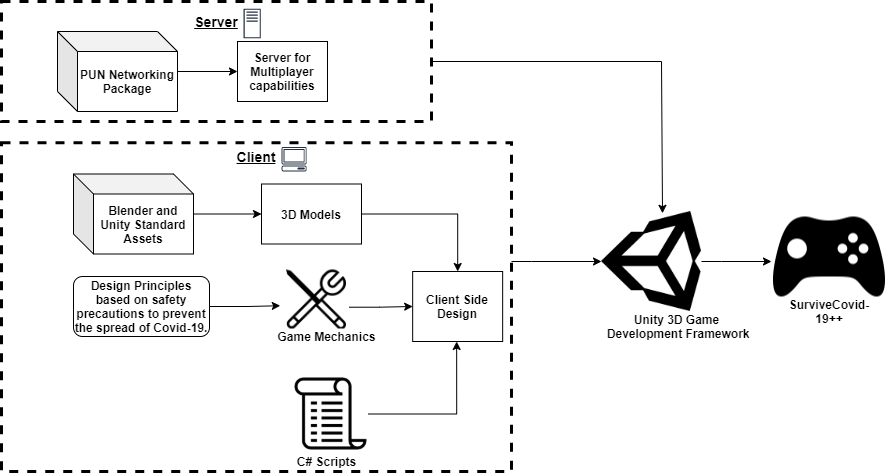}
    \caption{Development Approach of \textit{SurviveCovid-19++}}
    \label{fig:dev}
\end{figure}
\textit{SurviveCovid-19++} is a desktop based educational game, designed based on specific rules corresponding to players' actions. It thus, does not include any artificial intelligence agent in the current version. Hence, the features \textit{Artificial Intelligence}, that corresponds to inclusion of AI, as specified in \cite{buchinger2018guidelines} has not been considered in the design of \textit{SurviveCovid-19++}.  

\textit{SurviveCovid-19++} has thus been designed keeping all the relevant guidelines in view to facilitate in educating individuals about safety measures, vaccination and roles of different professions in the context of \textit{Covid-19}. The current versions consists of four stages, as shown in Fig. \ref{fig:map} with increasing difficulty as the stages progress.

\section{Development of \textit{SurviveCovid-19++}}
The development of SurviveCovid-19++ has been done using the Unity 3D Game Development Framework\footnote{\url{https://unity.com/}}. There are two major components to the game, \textit{Server} component and the \textit{Client} component as seen in Fig. \ref{fig:dev}. Development of the server component has been done with the help of \textit{Photon Networking Package} provided by \textit{Unity}. It is directly installed into the \textit{Unity Editor} for usage. This component is responsible for all the necessary multiplayer capabilities required for the game such as the synchronization of the game state and multiplayer room creation features. Combining the Photon Networking package with C\# scripting language allowed us to implement in-game chat functionality for the players. Using a \textit{Network Manager} script on client side we facilitate the connection of multiple Clients to the Server component over the internet.

The \textit{Client} component consists of all the graphical and mechanical aspects of the game. The graphical assets used in the game have been developed using the \textit{Blender 3D modelling software} and \textit{Unity Standard assets} pack. We proceeded to translate the safety guidelines for preventing the spread of \textit{Covid-19} virus, into specific game mechanics. For example, the completion of the game is made to be dependent upon the completion of individual goals specific to the different avatars along with a primary goal of collecting vaccine parts. If any individual player is not able to continue due to reduced health, the game is over for all the players. These mechanisms helped us in enforcing collaboration amongst different players. All the mechanisms were implemented using the C\# programming environment provided by the \textit{Unity} editor. All the quantitative elements necessary for keeping track of the game such as score, percentage completion of the goals, and health remaining are displayed to the user using UI elements. Finally, using the windows build settings from Unity, we build the \textit{exe} file for the game as shown in Fig. \ref{fig:dev}. 

% All statistical information regarding the state of the game and individual players is maintained using the Game Manager script. 

\section{User Scenario}
\begin{figure}
    \centering
    \includegraphics[width = \linewidth]{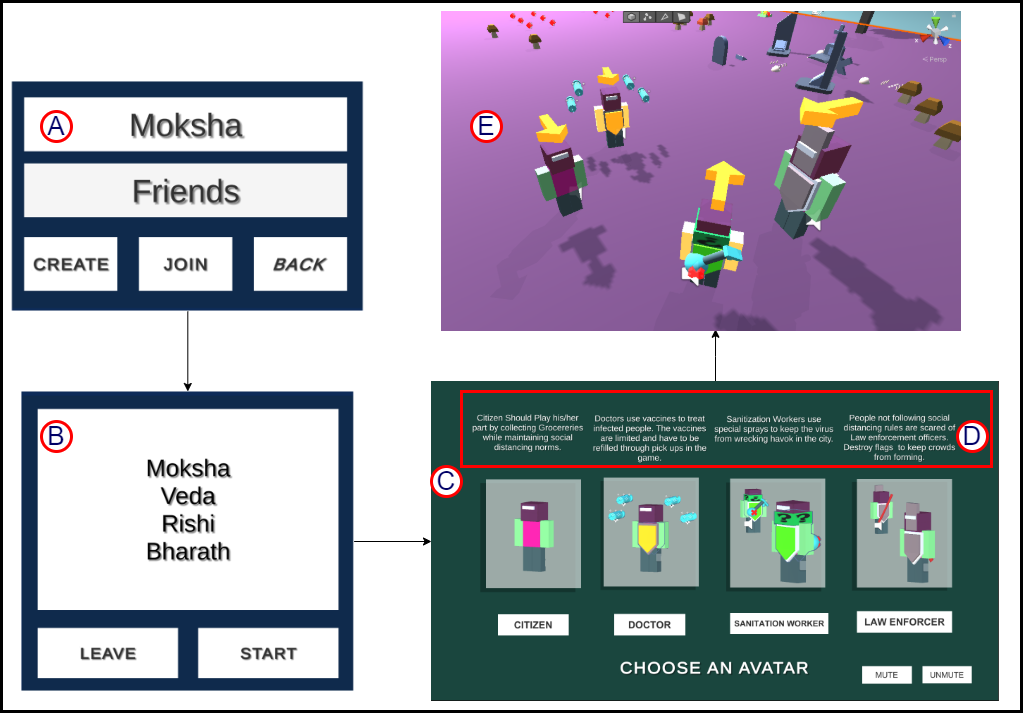}
    \caption{Snapshots of the game representing [A] User screen to enter player name and room name, [B] List of players who joined the room, with options for room creator to start the game, [C] Prompt screen to all users to select desired avatar, [D] List of instructions for each avatar, indicating the tasks to be done by the avatar and [E] Representing all four players at the beginning of the game.}
    \label{fig:sc1}
\end{figure}

\begin{figure}
    \centering
    \includegraphics[width = \linewidth]{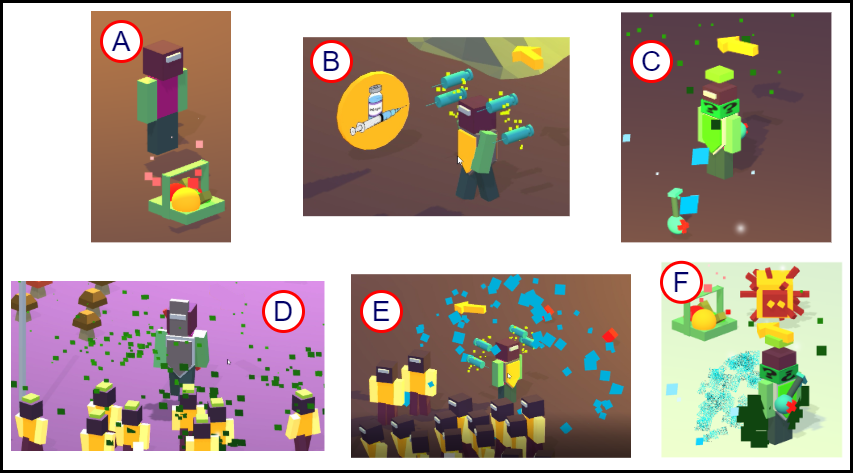}
    \caption{Snapshots of the game representing [A] Citizen collecting groceries, [B] Doctor collecting medicine refill, [C] Sanitation worker collecting disinfectant refill, [D] Law Enforcer clearing crowds, [E] Doctor treating affected people and [F] Sanitation worker sanitizing the area, and consequently killing the virus.}
    \label{fig:sc2}
\end{figure}

\begin{figure}
    \centering
    \includegraphics[width = \linewidth]{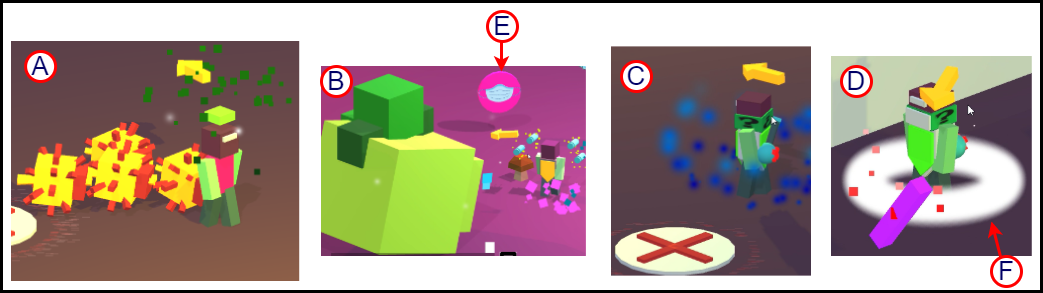}
    \caption{Snapshots of the game representing [A] Virus attacking citizen, [B] Stronger strain of the virus attempting to attack Doctor, [C] Sanitation worker leaving from hospital after being cured, [D] Sanitation worker collecting vaccine parts, [E] Masks to be collected by players when necessary, and [F] sanitation worker guarded by a ring on due collection of mask.}
    \label{fig:sc3}
\end{figure}

Consider \textit{Moksha} and her three friends from school, came across an article about collaborative efforts by people against \textit{Covid-19}. However, they were not able to understand the significance of it to fight against \textit{Covid-19}. Together they considered playing \textit{SurviveCovid-19++} game to gain deeper understanding about different collaborative efforts taken by people to tackle the \textit{Covid-19} virus. They download the game on their individual desktops and start the game. The players are now prompted to choose a display name. \textit{Moksha} creates a room as shown in Fig. \ref{fig:sc1}[A] and the rest of the members join using the same room name as depicted in Fig. \ref{fig:sc1}[B]. The players are then shown a player selection screen with 4 different avatars and their respective tasks as shown in Fig. \ref{fig:sc1}[C]. The players decide amongst themselves and choose their respective avatars, and start the game as shown in Fig. \ref{fig:sc1}[D]. 

\textit{Moksha} selects the \textit{Doctor} whose role is to heal infected people to avoid the spread of \textit{Covid-19}. Similarly her friend \textit{Veda} opts for the \textit{sanitation worker} whose role is to keep the surroundings safe by destroying viruses. \textit{Rishi} picks the role of the \textit{Citizen} who collects the essentials in the form of groceries. \textit{Bharath} picks the role of \textit{Law Enforcer} who makes sure that there are no large gatherings in order to prevent the virus outbreak.

Both doctor and sanitation worker  avatars have limited shots of vaccine and sanitizer respectively, which can be refilled by picking up appropriate pickups spread throughout the game as displayed in Fig. \ref{fig:sc2}[B] and Fig. \ref{fig:sc2}[C].  By using masks and sanitizers the citizen could navigate the map along with other avatars, while collecting essentials, as shown in Fig. \ref{fig:sc2}[A]. Law enforcer could break crowds as shown in Fig. \ref{fig:sc2}[D], which allows the player to quickly clear out large spaces. Doctor treats the infected people in the city as shown in Fig. \ref{fig:sc2}[E] and the sanitation worker uses disinfectant to clean the city and kill the virus as shown in Fig. \ref{fig:sc2}[F].

The four players are quickly overwhelmed by the waves of virus themed monsters which follow them and infect them upon contact, as depicted in Fig. \ref{fig:sc3} [A]. As they proceed through the stages, they encounter different varieties of viruses depicting different strains of \textit{Covid-19} as shown in Fig. \ref{fig:sc3}[B]. The four friends use the in-game chat functionality to interact with each other and come up with different strategies to finish the game. Collecting Mask and Sanitization pickups, as indicated by Figure\ref{fig:sc3}[E], provided the players with temporary protection from the virus as depicted in Fig. \ref{fig:sc3}[F]. Fig. \ref{fig:sc3}[C] depicts player visiting healthcare camps across the map to heal themselves when required, and getting healed. By collecting the primary vaccine parts, depicted in Fig. \ref{fig:sc3}[D], and completing individual goals the players proceed through the multiple stages together and are able to finish the game, all the while carefully following safe guidelines such as social distancing and usage of masks and sanitizers. Upon successful completion of the game the friends realize that \textit{SurviveCovid-19++} depicts how it would be very difficult to survive without all of the different avatars working together and fulfilling their roles to fight \textit{Covid-19}.

\section{Evaluation}
\subsection{Methodology}
\textit{SurviveCovid-19++} has been developed as a collaborative multiplayer healthcare educational game with an aim to facilitate better understanding about safety measures to be taken against \textit{Covid-19}, importance of roles played by varied professions and about importance and usefulness of taking vaccines. Better usability and player experience of game users determines reachability of the game. Hence, we evaluate \textit{SurviveCovid-19++} to assess the influence on players' understanding of individual and collaborative efforts and safety measures towards controlling the spread of \textit{Covid-19},  player experience,  and usability.

Similar educational games have been evaluated against similar factors such as player experience, usability, usefulness and so on \cite{santos2019risking, tsopra2020antibiogame}. Models such as MEEGA+ \cite{petri2016meega+}, Technology Acceptance Model (TAM) \cite{venkatesh2008technology}, Unified Theory of Acceptance and Use of Technology (UTAUT) \cite{venkatesh2011just} and so on have been used to evaluate player experience, usefulness, acceptance of the game among target audience and various other factors. Considering the relevance of MEEGA+, as it focuses on player experience and usability factors towards evaluating games, and the specificity of MEEGA+ to evaluate educational games, we evaluate \textit{SurviveCovid-19++} with an adapted version of MEEGA+ model. The player experience dimension presented in MEEGA+ model consists of nine factors which also include \textit{perceived learning (PL)} and \textit{user error protection (UEP)}. As \textit{SurviveCovid-19++} is not a classroom or course based game, comparing learning through the game with classroom-based learning is out of the scope of this evaluation. Moreover, \textit{SurviveCovid-19++} does not include any questionnaire that could lead users to make errors, indicating UEP factor also to be out of the scope of evaluation. Thus, we adapt MEEGA+ questionnaire to include only seven factors under player experience dimension.
\begin{table}[]

\caption{Questionnaire 1, used for Qualitative User Survey}
\label{tab:q1}
\begin{tabular}{|l|}
\hline
\textbf{Open-ended Questions }                                                            \\ \hline
1. What do you think is the role of doctors in controlling Covid-19?              \\ \hline
2. What do you think is the role of law enforces in controlling Covid-19?         \\ \hline
3. What do you think is the role of sanitization workers in controlling Covid-19? \\ \hline
4. What do you think is the role of common people in controlling Covid-19?        \\ \hline
5. Who do you think should work more towards controlling Covid-19? and why?      \\ \hline
6. Do you think controlling Covid-19 requires collaborative efforts?             \\ \hline
\end{tabular}
\end{table}

\begin{table}[]

\caption{Questionnaire 2 used for Quantitative User Survey and Corresponding Mean and Standard Deviation (SD) values}
\label{tab:q2}
\begin{tabular}{|l|l|l|l|}
\hline
\begin{tabular}[c]{@{}l@{}}\textbf{Quality} \\ \textbf{factor}\end{tabular} & \textbf{Questions}                                                                                                                                            & \textbf{Mean} & \textbf{SD}   \\ \hline
ABC                                                       & The theme of the game influenced my actions in real-time                                                                                             & 3.92 & 0.46 \\ \hline
ABC                                                       & \begin{tabular}[c]{@{}l@{}}After playing the game, I intended to use sanitizers more frequently\\  in my day-to-day life.\end{tabular}               & 3.92 & 0.53 \\ \hline
ABC                                                       & \begin{tabular}[c]{@{}l@{}}After playing the game, I intended to follow social distancing \\ more often in my day-to-day life\end{tabular}           & 3.96 & 0.69 \\ \hline
ABC                                                       & \begin{tabular}[c]{@{}l@{}}After playing the game, I intended to more frequently use \\ masks in my day-to-day life\end{tabular}                     & 4.21    & 0.56  \\ \hline
ABC                                                       & \begin{tabular}[c]{@{}l@{}}After playing the game, I intended to visit a doctor immediately \\ when I find symptoms of Covid-19\end{tabular}         & 4.07 & 0.85 \\ \hline
ABC                                                       & I was able to empathize with all professions after playing the game                                                                                  & 4.50 & 0.50  \\ \hline
LO                                                        & I understood the importance of taking vaccines in controlling Covid-19                                                                               & 4.35 & 0.67 \\ \hline
LO                                                        & \begin{tabular}[c]{@{}l@{}}I understood the importance of increased safety measures to be\\  taken for newer strains of Covid-19\end{tabular}        & 4.25    & 0.67    \\ \hline
LO                                                        & I could perceive that stronger strains need increased efforts to be controlled                                                                       & 4.50 & 0.57    \\ \hline
LO                                                        & Do you agree that collaborative efforts are required to control Covid-19?                                                                            & 4.78 & 0.41 \\ \hline
A                                                         & The game design is attractive (interface, graphics, cards, boards, etc.).                                                                            & 4.35    & 0.55    \\ \hline
Acc                                                       & The fonts (size and style) used in the game are easy to read.                                                                                        & 4.39 & 0.56 \\ \hline
Acc                                                       & The visual representation of  avatars used in the game is meaningful                                                                                 & 3.64    & 1.09 \\ \hline
L                                                         & I think that most people would learn to play this game very quickly.                                                                                 & 4.03 & 0.57 \\ \hline
L                                                         & Learning to play this game was easy for me.                                                                                                          & 3.92  & 0.71 \\ \hline
O                                                         & The game rules are clear and easy to understand.                                                                                                     & 4.39 & 0.62 \\ \hline
O                                                         & I think that the game is easy to play.                                                                                                               & 3.67 & 0.90  \\ \hline
C                                                         & \begin{tabular}[c]{@{}l@{}}When I first looked at the game, I had the impression that \\ it would be easy for me.\end{tabular}                       & 3.85  & 0.97  \\ \hline
Ch                                                        & \begin{tabular}[c]{@{}l@{}}The game does not become monotonous as it progresses \\ (repetitive or boring tasks).\end{tabular}                        & 4.25 & 0.58 \\ \hline
Ch                                                        & \begin{tabular}[c]{@{}l@{}}The game provides new challenges (offers new obstacles, \\ situations or variations) at an appropriate pace.\end{tabular} & 4.39    & 0.56 \\ \hline
Ch                                                        & This game is appropriately challenging for me.                                                                                                       & 4.14 & 0.59 \\ \hline
F                                                         & \begin{tabular}[c]{@{}l@{}}Something happened during the game (game elements, \\ competition, etc.) which made me smile.\end{tabular}                & 3.71 & 1.04 \\ \hline
F                                                         & I had fun with the game.                                                                                                                             & 4.32 & 0.54 \\ \hline
FA                                                        & \begin{tabular}[c]{@{}l@{}}There was something interesting at the beginning\\  of the game that captured my attention.\end{tabular}                  & 3.53 & 0.96  \\ \hline
R                                                         & The game contents are relevant to my interests.                                                                                                      & 3.46 & 0.83 \\ \hline
S                                                         & I would recommend this game to my colleagues.                                                                                                        & 4.35    & 0.55 \\ \hline
S                                                         & I feel satisfied with the things that I learned from the game.                                                                                       & 3.85 & 0.75  \\ \hline
S                                                         & It is due to my personal effort that I managed to advance in the game.                                                                               & 3.71 & 0.76 \\ \hline
S                                                         & \begin{tabular}[c]{@{}l@{}}Completing the game tasks gave me a satisfying \\ feeling of accomplishment.\end{tabular}                                 & 4.14 & 0.65 \\ \hline
SI                                                        & I  was able to interact with other players during the game.                                                                                          & 4.60 & 0.49 \\ \hline
SI                                                        & The game promotes collaboration among the players.                                                                                                   & 4.75 & 0.44 \\ \hline
SI                                                        & I felt good interacting with other players during the game.                                                                                          & 4.07 & 0.81 \\ \hline
\end{tabular}
\end{table}
Apart from the player experience aspect, pedagogical value of collaborative games has also been evaluated through questionnaires that assess player reaction, learning outcomes and behavioural/attitudinal changes \cite{cooney2020fun}. While the player reaction factor could be related to player experience factor in MEEGA+, the learning outcomes and behavioural/attitudinal changes correspond to influence of the game on the players. 

As we also wish to assess influence of \textit{SurviveCovid-19++} on the players, we adapt a mixed methods approach that includes an adapted MEEGA+ questionnaire, accompanied by factors evaluating pedagogical value, presented in \cite{cooney2020fun}. The factors considered for evaluation of the game are presented below.
\begin{itemize}
    \item Pedagogy-related factors
    \begin{itemize}
    \item Learning Outcome (LO)
    \item Attitudinal/Behavioural Change (ABC)
    \end{itemize}
    
    \item Player Experience
    \begin{itemize}
        \item Focused Attention(FA)
        \item Fun (F)
        \item Challenge (Ch)
        \item Social Interaction (SI)
        \item Confidence (C)
        \item Relevance (R)
        \item Satisfaction (S)
    \end{itemize}
    
    \item Usability
    \begin{itemize}
        \item Learnability (L)
        \item Operability (O)
        \item Aesthetics (A)
        \item Accessibility (Acc)
    \end{itemize}
    
\end{itemize}

% Please add the following required packages to your document preamble:
% \usepackage[normalem]{ulem}
% \useunder{\uline}{\ul}{}

\subsection{User Study}
We evaluated \textit{SurviveCovid-19++} based on the factors discussed above with seven teams, comprising of four players each.
All the players were requested to answer a questionnaire comprising of five open-ended and one multiple choice question, that correspond to understanding of the player, with respect to roles played by individuals of varied professions and player perception about need for collaboration to control \textit{Covid-19} (listed in Table \ref{tab:q1}). After answering the questionnaire, each team played the game for fifteen minutes. One player in each team was asked to create a group and the rest of the players of the team were requested to join that team. Each player in the team were requested to select an avatar of their choice and start playing the game. Each team played the game for at least 1 time in a span of fifteen minutes. After playing the game, the participants were requested to answer two questionnaires. One questionnaire (Questionnaire 1, presented in Table \ref{tab:q1}) consisted of the same questions that were asked before playing the game, to understand perception of the players, and another five-point likert-scale based questionnaire (Questionnaire 2, presented in Table \ref{tab:q2}) consisted of questions corresponding to the player experience, usability and influence dimensions. The Questionnaire 2 consisted of 36 questions which included four questions corresponding to player demographics and 32 questions corresponding to quality factors of the game.

\section{Results}
Among the 28 volunteers who participated in the user survey, 24 of them were male and four of them were female. Most of the participants are undergraduate students and have agreed that they play games frequently. Based on the answers given by the participants for the Questionnaire 1 before playing the game, we observe that about six of the participants did not correctly interpret the role of citizens and law enforcers in controlling the pandemic. Three of the participants wrongly interpreted the role of sanitation workers and about 60\% (17 out of 28) of the participants have misinterpreted the need for collaborative efforts towards controlling the pandemic. When all the participants answered Questionnaire 1 after playing the game, we observed change in their interpretation. All the participants described the roles of individuals with different professions correctly to a reasonable extent and also recognized the need for collaborative efforts to control the pandemic. This implies that \textit{SurviveCovid-19++} has contributed to better understanding of the players in the context of \textit{Covid-19}.
\begin{figure}
    \centering
    \includegraphics[width = \linewidth]{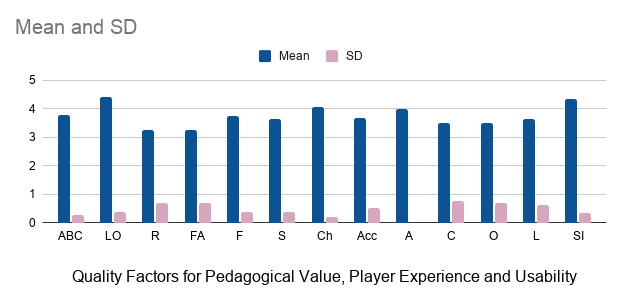}
    \caption{Results of Quantitative User Survey in terms of Mean and Standard Deviation of Quality factors}
    \label{fig:msd}
\end{figure}

The mean and standard deviation for each question in Questionnaire 2 are presented in Table \ref{tab:q2}. The mean value greater than 3.46 for all the questions in Questionnaire 2 indicates that users had a reasonably positive experience with the game. All the ten questions corresponding to pedagogy-related factors (LO and ABC) have mean greater than or equal to 3.92, indicating that \textit{SurviveCovid-19++} could successfully teach need for collaborative efforts and role of each individual in the community, and could successfully influence the players to follow safety measures towards controlling \textit{Covid-19}. 

The mean and standard deviation values of all the thirteen quality factors have been plotted in Fig. \ref{fig:msd}. 
Mean value greater than 3.46 for all of the quality factors, with low standard deviation, indicates that majority of the players agreed that \textit{SurviveCovid-19++} is useful, interesting and usable.

We calculate the quality score of \textit{SurviveCovid-19++} based on MEEGA+ model by considering the mean values for factors under the dimensions - player experience and usability. 
Corresponding Cronbach Alpha values presented in MEEGA+ for each dimension are applied to the means of player experience and usability factors, and the resultant value is normalized by multiplying it by 10. This resulted in a quality score of 71.90. According to the MEEGA+ evaluation model, games with quality score greater than 65 are considered to be of excellent quality, and that less than 42.5 are considered to be poor quality. The game has a quality score of 71.90, indicating inclusion of the game in excellent category. This result indicates that such games are challenging for learners, without monotonous activities and clear rules to play the game.

% The mean values for each of the quality factors were considered to calculate the quality score based on MEEGA+ model. Applying the Cronbach Alpha value for means of player experience (FA, F, Ch, S, R) and usability (L, O, As, Acc) factors, and then normalizing the resultant value by multiplying it by 10 resulted in quality score of 69.3. The MEEGA+ evaluation theory states that games with quality score greater than 65 are of excellent quality, while that between 42.5 to 65 are considered to be of better quality. The quality score of \textit{SurviveCovid-19} is 69.3, indicating that the game could be included in excellent quality category. This result indicates that such games are challenging for learners, without monotonous activities and clear rules to play the game.

\section{Discussion and Limitations}
\textit{SurviveCovid-19++} enforces players to collaboratively work towards controlling \textit{Covid-19}, by following the safety measures, collecting vaccine parts and other individual collectibles. Thus, it could educate players about importance of safety measures and vaccination, and role of individuals in different professions towards controlling the spread of the virus. This game can be played by wider range of audience, without any constraint on educational background, except for basic understanding of English language. The only hardware constraint is that it can be played only on desktops with internet connection. 

The current multiplayer version accommodates only four players per room, analogous to the four avatars being considered in the game and has only four stages. More number of players and stages can be accommodated, with minimal technical modifications in the game. The current version requires the game application to be downloaded on the players' local machine. It does not support playing the game on a browser. However, it can be extended as a browser-based game with minimal efforts. \textit{SurviveCovid-19++} currently supports only desktops/laptops and is not compatible with mobile devices. Also, apart from the instructions presented in the chat box, \textit{SurviveCovid-19++} does not support any other in-game AI/non-AI based assistant or in-game dialogues, to guide the players. However, this being a collaborative game that supports chat application, players could communicate with each other and play the role of AI assistant by conveying human suggestions and instructions.

As we focused on influencing behaviour of the players towards following safety measures for \textit{Covid-19}, and collaborating towards controlling the spread of the pandemic, we performed an in-depth evaluation with 28 players and closely observed their behavioural changes when they encounter various scenarios in the game, and assessed their experience with the game. However, we plan to extend the evaluation with a larger number of players.

\section{Conclusion and Future Work}
In this paper, we presented \textit{SurviveCovid-19++}, a multiplayer, collaborative desktop game, aimed to facilitate learning about vaccines, safety measures to be followed, roles played by different professions (\textit{doctor}, \textit{citizen}, \textit{law enforcer} and \textit{sanitation worker}) and need for collaborative effort to control the spread of \textit{Covid-19}. It has been designed based on the design principles specified for collaborative games and for games in the healthcare context to ensure better reachability and value of the game. 
\textit{SurviveCovid-19++} has been evaluated to assess player experience, usability and its pedagogical value, through a user survey based on an adapted MEEGA+ questionnaire, with seven teams of four players each. The evaluation comprised of a pre and post test qualitative questionnaire to assess change in perception of the users and a likert-scale based quantitative questionnaire to assess player experience, usability and influence of the game on players. The results of the evaluation were positive, with observed attitudinal changes and learning outcomes among all the participants, and all the participants willing to recommend \textit{SurviveCovid-19++} to their colleagues.

% Games are capable of motivating users to learn and also increase retention capacity of the players. This feature of games has resulted in increased adaptation of games in the educational context. They are also used to treat patients, to educate individuals about healthy habits, and also to bring awareness about certain diseases, in the healthcare context. Collaborative educational games are capable of increasing teamspirit and in encouraging to learn newer concepts. Considering the usefulness of games in healthcare context and need to teach the importance of collaborative efforts, and need to follow safety precautions towards controlling \textit{Covid-19}, collaborative games could be helpful in the context of \textit{Covid-19}. 

% Though \textit{SurviveCovid-19} exists in the literature, which aims to bring awareness about safety measures to be followed against \textit{Covid-19}, it is a single player game and does not include any information about vaccines, different strains of the virus, roles of different professions in controlling the virus and so on.  Hence, we present \textit{SurviveCovid-19++}, as a collaborative multi-player game that aims to educate users about safety precautions to be followed, different strains of the virus, importance of vaccines and role of individuals with different professions. 

\textit{SurviveCovid-19++} currently supports only four avatars, corresponding to four different professions. More number of avatars could be added in the game, to support wide range of professions in-game, in the future versions. The current multi-player version only supports up to four players, which could further be extended in the future. As a part of future work, we also plan to facilitate users to create and customize their own avatars. Also, \textit{SurviveCovid-19++} currently supports only four stages. We plan to increase the number of stages by adapting a procedural approach to generate stages, and increase difficulty in each stage. Moreover, the current version of the game is supported only on desktop, which can be extended to support mobile devices as well. 

\begin{acks}
We would like to thank the volunteers for their valuable and honest feedback and The Richard Lounsbery Foundation for funding and encouraging this project.
\end{acks}

%%
%% The next two lines define the bibliography style to be used, and
%% the bibliography file.
\bibliographystyle{ACM-Reference-Format}
\bibliography{sample-base}

%%
%% If your work has an appendix, this is the place to put it.

\end{document}